\documentclass[aps,prl,superscriptaddress,twocolumn,showpacs]{revtex4}
\usepackage[pdftex]{color,graphicx}
\usepackage{amssymb,amsmath}

\begin{document}

\title{Whispering-gallery mode micro-kylix resonators}
\author{Mher Ghulinyan} 
\affiliation{Advanced Photonics \& Photovoltaics, Fondazione Bruno
Kessler - irst, via Sommarive 18, Povo, Trento, Italy}
\author{Alessandro Pitanti}
\affiliation{Nanoscience Laboratory, Dept. Physics, University of
Trento, Via Sommarive 14, Povo, I-38050 Trento, Italy}
\author{Georg Pucker}
\affiliation{Advanced Photonics \& Photovoltaics, Fondazione Bruno
Kessler - irst, via Sommarive 18, Povo, Trento, Italy}
\author{Lorenzo Pavesi}
\affiliation{Nanoscience Laboratory, Dept. Physics, University of
Trento, Via Sommarive 14, Povo, I-38050 Trento, Italy}

\begin{abstract}
Owing to their ability to confine electromagnetic energy in
ultrasmall dielectric volumes, micro-disk, ring and toroid
resonators hold interest for both specific applications and
fundamental investigations. Generally, contributions from various
loss channels within these devices lead to limited spectral windows
(Q-bands) where highest mode Q-factors manifest. Here we describe a
strategy for tuning Q-bands using a new class of micro-resonators,
named micro-kylix resonators, in which engineered stress within an
initially flat disk results in either concave or convex devices. To
shift the Q-band by $60$nm towards short wavelengths in flat
micro-disks a 50\% diameter reduction is required, which causes
severe radiative losses suppressing Q's. With a micro-kylix, we
achieve similar tuning and even higher Q's by two orders of
magnitude smaller diameter modification (0.4\%). The phenomenon
relies on geometry-induced smart interplay between modified
dispersions of material absorption and radiative loss-related
Q-factors. Micro-kylix devices can provide new functionalities and
novel technological solutions for photonics and micro-resonator
physics.
\end{abstract}

\pacs{42.60.Da, 42.70.Qs, 78.67.Bf}

\maketitle

Recent technological advances in micro- and nanophotonics have
boosted the realization and testing of extremely high Q-factor
optical resonator systems. Much interest is focused on
two-dimensional whispering-gallery mode (WGM) resonators
\cite{rayleigh,ilchenko}, such as micro-disks \cite{mudisks}, rings
\cite{murings} and toroids \cite{toroid}; these are chip-integrable
and offer a wide spectrum of possible applications ranging from
microdisk ($\mu$-disk) lasers \cite{lasers1,lasers2} to sensing
\cite{deio}. Besides, WGM resonators are interesting for fundamental
physics, such as cavity quantum electrodynamics \cite{coptomech},
individual atom-photon quantum interactions \cite{Dayan} and for
testing fascinating optical phenomena like radiation pressure forces
\cite{forces}.

The main figure of merit of a $\mu$-disk is the $Q$-factor, which is
a measure of the energy stored in the resonator versus the energy
dissipated per round trip. $Q$-factor engineering is getting
relevance since it improves significantly device performances. In
addition, for active $\mu$-disks high $Q$'s are desirable in
spectral windows of maximum material gain; possible resonant mode
competition can be overcome with an appropriate tuning of spectral
positions of high-Q modes towards the gain band peak, without
modifying the free-spectral range (FSR) of resonator modes. On the
other hand, when WGM detection schemes are used in biosensing
applications, analyte molecules spoil cavity $Q$-factor by binding
to the resonator surface. In this case, an enhanced detection
sensitivity would benefit from high-$Q$'s at shorter wavelengths,
where analyte molecules possess stronger absorption. Basic schemes
to shift high $Q$-factors from long to short wavelengths rely on
achieving a stronger confinement of WGM modes within the resonator
device; this is achieved either by increasing the $\mu$-disk
diameter or the thickness, both imposing larger device dimensions.
As a consequence, these solutions cause unavoidable reduction of
mode FSR and appearance of higher order mode families.

\begin{figure*}
\centering\includegraphics[width=14cm]{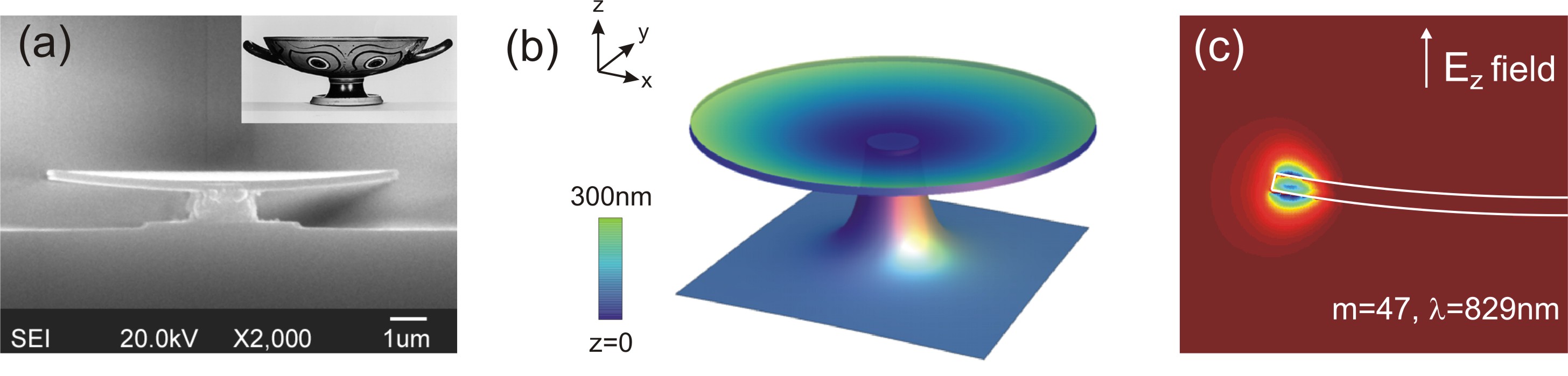} \caption{ The
micro-kylix resonator. (a) Cross-sectional scanning electron
micrograph (SEM) of a micro-kylix resonator. Inset: The Greek
wine-drinking cup - $\kappa\upsilon\lambda\iota\xi$, with a broad
relatively shallow body raised on a stem from a foot \cite{kylix}.
(b) Three-dimensional view of an analytically calculated bent-up
resonator using experimental dimensions (experimental bending radius
of 47~$\mu$m). The extremes of the scale-bar represent the height
difference between the edge and the center of a 200~nm-thick disk.
(c) Electrical field distribution of the TM-polarized
whispering-gallery mode shows a rather good confinement within the
core of the $\mu$-kylix resonator} \label{kylixsem}
\end{figure*}

Here we report on a new class of $\mu$-disk resonators, which we
call \emph{micro-kylix} ($\mu$-kylix) resonators
(Fig.~\ref{kylixsem}(a)), in analogy with the Greek wine-drinking
cup - $\kappa\upsilon\lambda\iota\xi$, (inset photo,
Fig.~\ref{kylixsem}(a)) \cite{kylix}. These present a bent
$\mu$-disk configuration, which, owing to an appropriate
stress-engineering approach, can be realized both as a bent-up
(cup-like, $\mu$-kylix) or bent-down (umbrella-like, inverse
$\mu$-kylix) resonators. These devices support perfectly
whispering-gallery modes and, at the same time, posses new and
unexpected physical characteristics. The most striking of these,
perhaps, is that their geometry induces a tuning of the $Q$-factor
band of resonator modes towards the stronger material absorption
range without degrading highest Q's. Importantly, this tuning scheme
does not require larger device sizes, but rather utilizes
self-adjustment properties of originally stressed resonator core.
Remarkably, the $\mu$-kylix resonator benefits from unmodified FSR
and cleaner WGM spectra due to the absence of higher order mode
families.

We used a finite-difference time-domain (FDTD) tool to check if a
$\mu$-kylix acts as WGM resonator. Simulations have been performed
using a freely-available FDTD software package \cite{harminv1},
which exploits the azimuthal symmetry of micro-disk resonator to
reduce the computational load. WGM resonances have been extracted
using a harmonic inversion algorithm included in the software
\cite{harminv2}. Starting with a 10~$\mu$m diameter flat
micro-resonator, we first verified that it supports WGMs, and, then,
we applied a bending to the flat disk in the $z$-direction
(Fig.~\ref{kylixsem}(b)). Indeed, the numerical simulation confirmed
that the $\mu$-kylix supports guided modes. As an example,
Fig.~\ref{kylixsem}(c) shows the electrical field distribution of a
TM-polarized mode (azimuthal mode number $m=47$, $\lambda=829~nm$)
close to the edge of the $\mu$-kylix resonator.

To realize the $\mu$-kylix we used a combination of two materials,
silicon-rich silicon oxide $SiO{_x}$ (SRO) and silicon nitride
$Si{_3}N{_4}$. We take advantage of the fact that these two
materials posses different thermal expansion coefficients, which
lead to unavoidable stresses at the end of the deposition process.
The stress can be engineered by accurate control of the deposition
temperature and the layer thickness. After disk definition and post
formation to isolate the disk from the substrate, the accumulated
stress gradients through the layer interfaces bend the micro-disk
device Fig.~\ref{kylixsem}(a)).

For $\mu$-kylix we first deposit a 160~nm-thick $SiO_x$ layer on top
of crystalline silicon wafers from a mixture of silane ($SiH_4$, 65
sccm) and nitrous oxide ($N_2O$, 973 sccm) gases using a
parallel-plate plasma enhanced chemical vapor deposition (PECVD)
chamber at T=300$^{\circ}C$  \cite{optexp}. Si-nc were formed
through a successive one hour annealing in an $N_2$ atmosphere at
T=1100$^{\circ}C$ \cite{pavesi3}. After the high-T treatment the SRO
layer densifies down to 110~nm due to the release of hydrogen and
micro-voids, present in the as-deposited layer. The residual
compressive stress on SRO/bulk-Si interface was measured to be
$-100$~MPa. In a next step, a similarly thick $Si_3N_4$ layer was
deposited at 780$^{\circ}C$ using a low pressure chemical vapor
deposition (LPCVD) technique (1.25~GPa tensile stress). A
photolithographical patterning of disk arrays was followed by dry
(anisotropic, disk formation) and wet etching (isotropic, post
formation) steps.

In this way, $\mu$-kylixes are obtained by engineering the stress in
the bilayer structure. The simplicity of such a technological
solution is combined to a large degree of freedom for not only the
choice of materials, but also control of both the degree and the
direction of disk reshaping. The degree of bend is strictly related
to the thickness ratio of the two different materials, while the
direction of bend (cup or umbrella) can be easily changed inverting
the materials. An example of a $\mu$-kylix is shown in the
cross-sectional scanning electron micrograph of
Fig.~\ref{kylixsem}(b). In our case we choose this set of materials
for their interest in silicon photonics \cite{pavesi1}. Optically
active $\mu$-kylix and $\mu$-disk resonators are easily realized by
annealing the SRO layer and, thus, inducing a phase separation of
SiO$_2$ and silicon with the formation of Si nanocrystals (Si-nc)
\cite{pavesi2}. These show strong room-temperature emission in a
wide spectral range (700-900 nm) \cite{pavesi3}, and a smooth
increasing absorption band with a threshold at about 800 nm
\cite{pavesi4}.

\begin{figure} [t!]
\centering\includegraphics[width=8.5 cm]{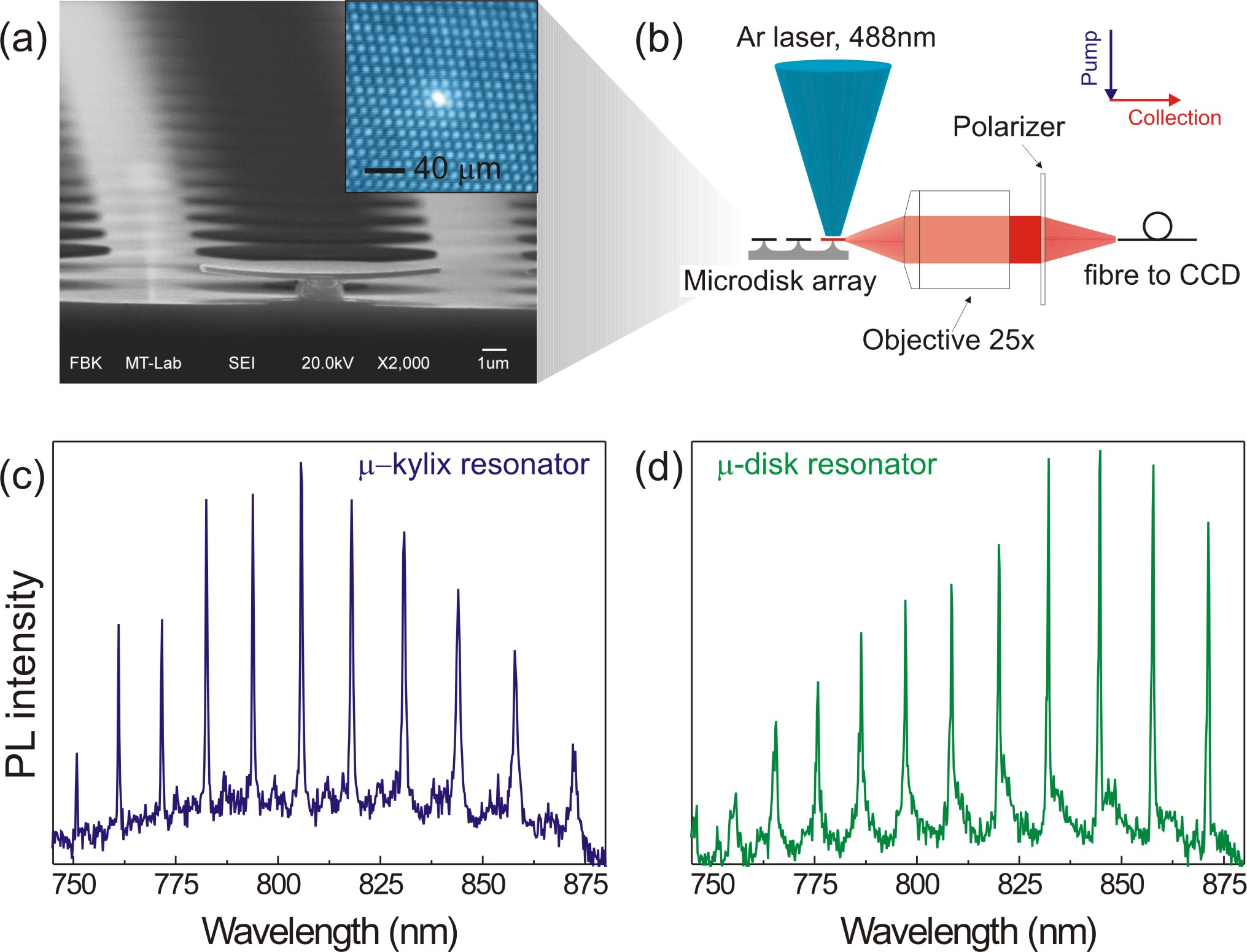} \caption{ WGM
emission from $\mu$-kylix and $\mu$-disk resonators. (a) An SEM
image of the $\mu$-kylix array with a cross-sectioned $\mu$-kylix
resonator on the sample edge. The inset shows an optical image of
$\mu$-kylix array, with the bright spot corresponding to the Si-nc
emission from a single device. (b) Micro-photoluminescence (PL)
setup used for testing single devices. (c) Measured emission
spectrum of the $\mu$-kylix resonator, showing sharp
whispering-gallery resonances of the lowest-order family raising out
from the broad PL band of Si-nc. (d) Measured emission spectrum of
the $\mu$-disk resonator. The 2nd-order family modes manifest as
broad and damped peaks next to intense 1st-order family resonances.}
\label{wgm}
\end{figure}

We measured the room-temperature micro-photoluminescence emission
from single resonators in a free-space collection configuration (see
Fig.~\ref{wgm}(a,b)). In our experiments we excited individual
micro-kylix resonators focusing with a long-working distance
objective the 488~nm line of a CW Argon laser and collected the
emission of Si nanocrystals from the edge of the sample through a
25x objective. The signal was then spatially filtered to reduce the
numerical aperture of the collecting set-up, selected by a polarizer
and sent to a monochromator interfaced with a cooled silicon charge
coupled device (CCD).

We observed sharp, sub-nanometer modal peaks raising out from the
wide emission band of Si-nc due to WGM. The measured spectrum
(Fig.~\ref{wgm}(c)), at first glance, is qualitatively similar to
that observed from single-layer-SRO $\mu$-disks \cite{optexp}, with
the difference that the narrowest resonances here manifest at
shorter wavelengths.

\begin{figure} [t!]
\centering\includegraphics[width=7.5 cm]{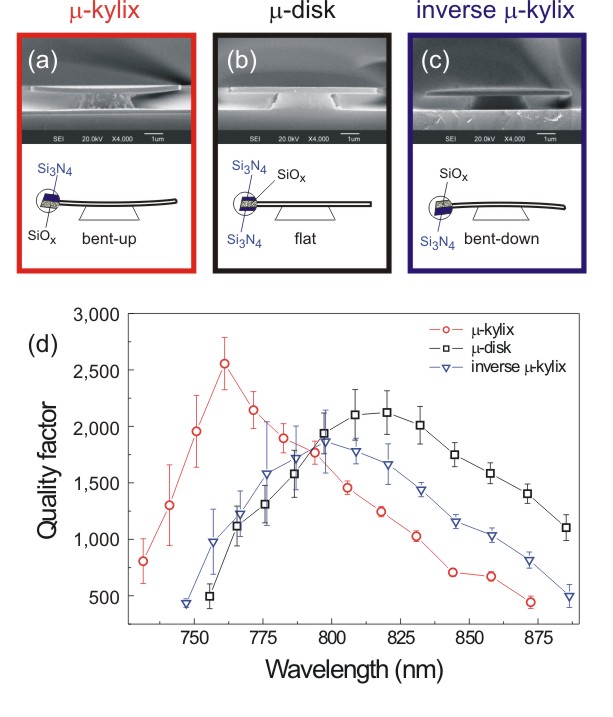}
\caption{Tuning of quality factors in $\mu$-kylix resonators.
Scanning electron micrographs and corresponding shape sketches for
(a) $\mu$-kylix, (b) $\mu$-disk and (c) inverse $\mu$-kylix
resonators, respectively. (d) Measured Q-factor bands for
differently shaped resonators.} \label{qshift}
\end{figure}

This observation led us to perform a second experiment. For this, we
prepared a flat $\mu$-disk, consisting of the same amount of SRO and
Si$_3$N$_4$ as in a $\mu$-kylix resonator (Fig.~\ref{qshift}(a)). In
order to have the disk flat, we sandwiched the SRO layer between two
55~nm-thick $Si_3N_4$ layers (Fig.~\ref{qshift}(b)). This way the
total thickness and the weighted-average refractive index of the
$\mu$-disk were similar to those of the $\mu$-kylix.

We measured the WGM emission spectrum of the $\mu$-disk
(Fig.~\ref{wgm}(d)) and compared with that of the $\mu$-kylix.
Lorentzian fit analysis of all appreciable peaks, plotted in
Figure~\ref{qshift}(d), evidences an important modification of
Q-factors in $\mu$-kylix. We observe that the Q-band maximum in the
$\mu$-kylix is blue-shifted by almost 60~nm with respect to that of
the $\mu$-disk. As a consequence, this shift results in higher
Q-factors (roughly four-fold) at wavelengths ($\sim760$~nm), where
our active material, SRO, has stronger absorption. Let us note that
the reported values for the highest Q-factors are limited by the
spectral resolution of our experimental setup; this does not
interfere with the main results of this work.

Before explaining these observations and revealing the physical
mechanisms behind, we first address the origin of spectral
dispersion of Q-factors. In $\mu$-disk resonators, the quality
factors for a certain radial family of resonances form a band, owing
to wavelength dispersions of Q's related to various loss channels.
In $\mu$-disks of few micrometers in size, the total Q-factor
dispersion is defined mainly by the interplay between the radiative
and material Q's, expressed as
\begin{equation}\label{qu}
    Q^{-1}_{total}(\lambda)\approx
    Q^{-1}_{rad}(\lambda)+Q^{-1}_{mat}(\lambda)+Q^{-1}_{i},
\end{equation}
with the semi-empirical term $Q_i^{-1}$ accounting for losses due to
nonidealities in real devices. Moreover, in active micro-resonators
the measured Q-band of WGM emission is limited to the spectral
region of the active material luminescence band, in our case, that
of Si-nc.

\begin{figure*}
\centering\includegraphics[width=12cm]{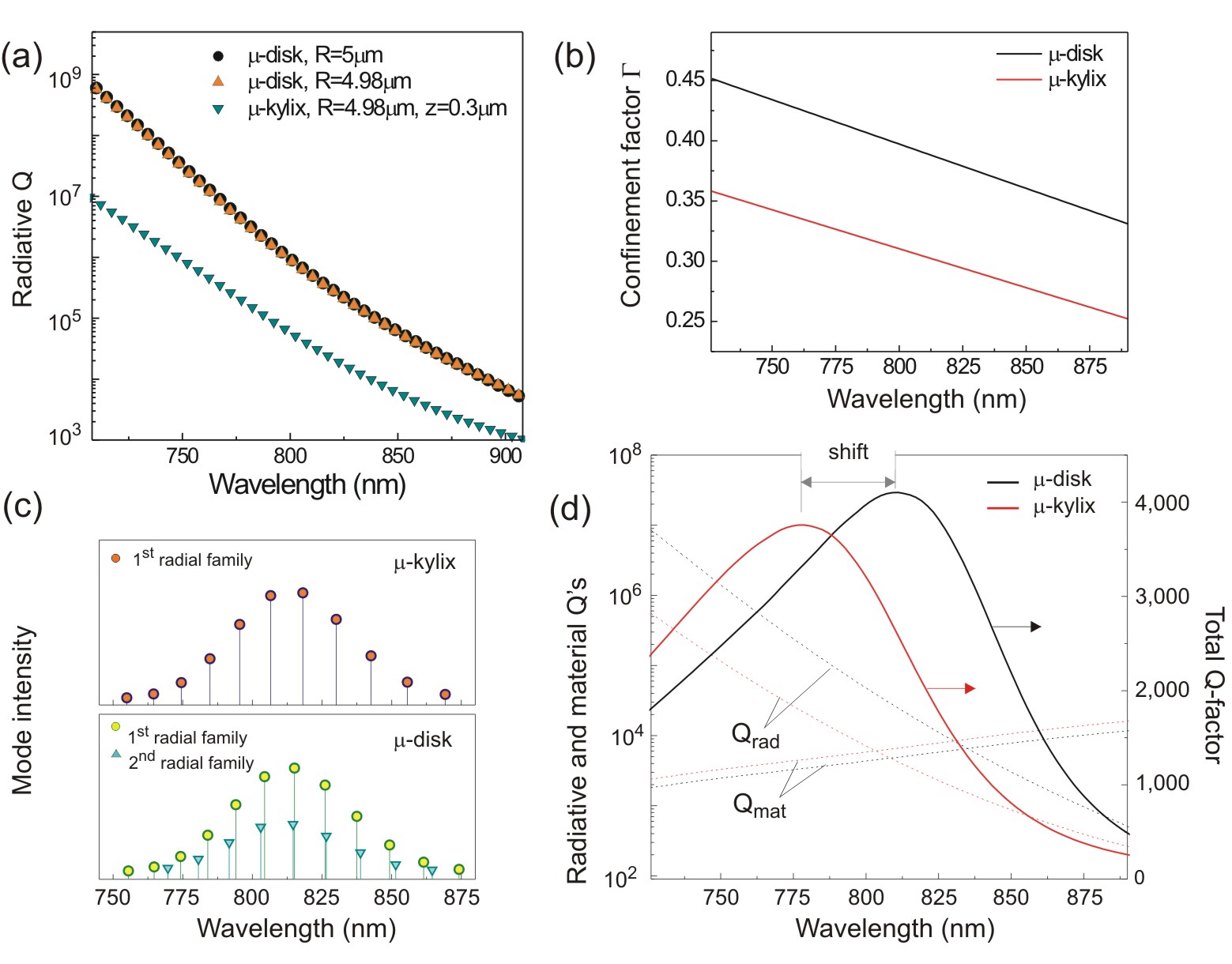} \caption{
Out-of-plane radiative losses and modified Q-dispersions. (a) Small
reduction of the disk diameter by only 0.4\% (from 10~$\mu$m to
9.96~$\mu$m) induces negligible change in radiative $Q$ for
$\mu$-disks ($\bullet$ and $\blacktriangle$, respectively). Instead,
bending the originally flat disk to an effective diameter of
9.96~$\mu$m (thus forming a $\mu$-kylix) a tenfold attenuation of
quality factor is observed. (b) Confinement factor dispersions,
calculated for different resonator configurations. (c) (top panel)
Numerical simulations of a non-absorbing $\mu$-kylix show that due
to a weaker mode confinement only 1st-order radial family manifests.
Bottom panel shows that $\mu$-disks support quite intense 2nd-order
families. (d) Numerically calculated dispersions of radiative,
material and total quality factors in a $\mu$-disk and a $\mu$-kylix
according to Eq.~\ref{qu}.} \label{radqs}
\end{figure*}

In our experiment, the possibility of having modified radiative
losses is rather straightforward; stress-induced formation of
$\mu$-kylix resonators from nominally $10~\mu$m flat disks, results
in a slightly smaller (by $\sim 50$~nm) external diameter. For
$\mu$-disks, such a small modification of the diameter (by 0.4\%
only), is not expected to affect significantly $Q_{rad}$. For
200~nm-thick passive flat devices of 10~$\mu$m and 9.96~$\mu$m of
diameter, typical $Q_{rad}$ values at $\lambda=800~nm$ are of the
order of $10^6$ and differ by only $\sim5\%$ (Fig.~\ref{radqs}(a)).
Numerical simulations, however, show that in a $\mu$-kylix resonator
of 9.96~$\mu$m diameter $Q_{rad}$ reduces by an order of magnitude
with respect to the original $\mu$-disk Fig.~\ref{radqs}(a),
down-triangled plot). A tenfold reduction in radiative $Q$
associated to disk reshaping is an interesting result which has
never been addressed. To model this, we introduce the concept of
out-of-plane bending loss ($1/Q_{bend}(z)$), which is a function of
the bending degree, i.e. of the coordinate $z$ of the out-of-plain
lifting of the disk edge. Thus, the radiative $Q$ for a $\mu$-kylix
can be modeled as
\begin{equation}\label{qurad}
    Q^{-1}_{rad}(R,z)=Q^{-1}_{rad}(R)+Q^{-1}_{bend}(z),
\end{equation}
where $Q_{rad}(R)$ is the radiative Q-factor for a flat resonator of
radius $R$. FDTD calculations show that most of the reduction of
Q$_{rad}$ are due to Q$_{bend}$ (Fig.~\ref{radqs}(a)).

The change in $Q_{rad}$ leads to an important spectral blue-shift of
the Q-band with an overall reduction of Q values. Since in
experiment we only observe a shift but not a reduction of Q's, we
conclude that in a $\mu$-kylix the Q$_{mat}$ is also affected
(increased). This scenario is surprising.

The material quality factor is generally defined as $Q_{mat}=2\pi
n_{g}/(\Gamma \alpha \lambda)$, where $n_{g}$ and $\Gamma$ are the
group index and optical confinement factor of the mode in the active
layer, respectively, and $\alpha$ is the material absorption at the
wavelength $\lambda$. The material absorption has been measured by
ellipsometry and was found to be the same for the different disk
configurations (the stress does not affect the absorption of
Si-nc's). Possible modifications in $Q_{mat}$, therefore, should
only result from differences in group indices and/or confinement
factors. We performed further numerical simulations and reveal very
similar $n_{g}$ for both $\mu$-disk and $\mu$-kylix
($n_{g}\approx$2.02 and 2.12 at 800~nm, respectively). While
slab-waveguide simulations for various layer sequences show almost
identical $\Gamma$-values, in a disk geometry important differences
are observed. Namely, the $\mu$-kylix shows smaller $\Gamma$-values
than the $\mu$-disk ($\Gamma\approx$0.31 and 0.4 at 800~nm,
respectively) (Fig.~\ref{radqs}(b)). Since $Q_{mat}$ is inversely
proportional to $\Gamma$, a roughly 25\% less confinement leads to
an appreciable change in material Q's. This, in its turn, modifies
the Q-band, shifting it further to shorter wavelengths, but, more
importantly, does not degrade highest Q values. Moreover, lower
$\Gamma$ values explain why, on the contrary to $\mu$-disks, the
$\mu$-kylix resonators do not support higher order mode families
(Fig.~\ref{wgm}(c,d)); the overall reduced confinement in
$\mu$-kylix alters guided mode propagation for higher order families
(see Fig.~\ref{radqs}(c)). Combining our numerical results for both
modifications in $Q_{rad}$ and $Q_{mat}$, we reproduce qualitatively
the experimentally observed Q-band tuning (Fig.~\ref{radqs}(d)).

\begin{figure} [t!]
\centering\includegraphics[width=7cm]{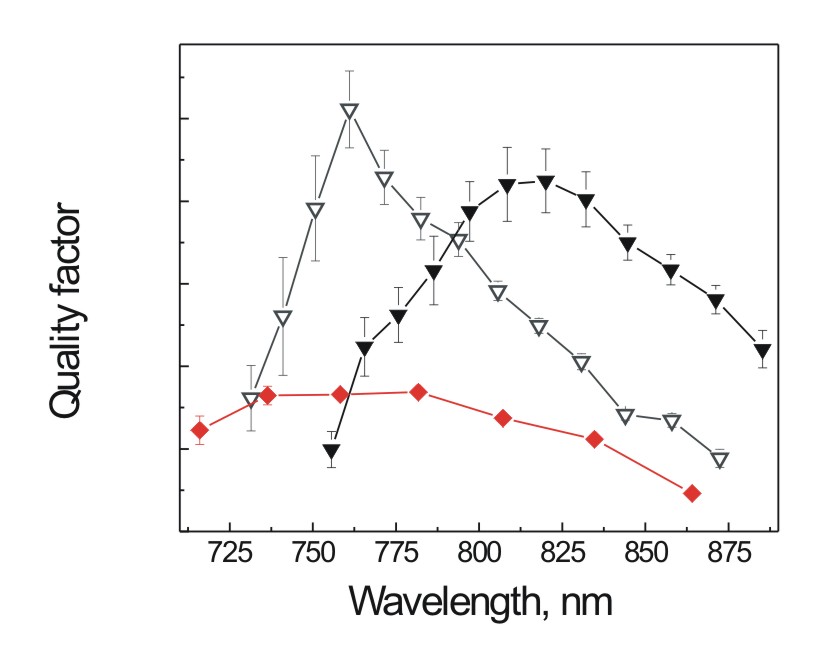} \caption{ Q-band
shift in smaller flat disks. A 60~nm spectral shift of the Q-band is
measured for twice smaller (5~$\mu$m) $\mu$-disks, while the same
shift can be achieved with a $\mu$-kylix of only 0.4\% smaller
effective diameter.} \label{shift2}
\end{figure}

For a final confirmation that the disk geometry is responsible for
the tuning of Q-factors, we performed a third experiment, in which
inverse $\mu$-kylixes were realized (Fig.~\ref{qshift}(c)) and
tested. The band shift for these devices is less pronounced
(Fig.~\ref{qshift}(d), blue empty circles), in accordance with the
minor stress of the disk; much lower temperature of the
$SiO{_x}$/$Si{_3}N{_4}$ interface formation ($T=300^{\circ}C$ in the
PECVD deposition chamber) and eventual high-temperature annealing
for Si-nc formation result in a reduced stress. Consequently, the
inverse $\mu$-kylix is characterized by a larger bending radius,
$\approx83 \mu$m (smaller $z$) and, hence, lower out-of-plane
losses, Q$_{bend}^{-1}$.

It is important to note, that in conventional $\mu$-disk resonators
smaller diameters blue-shift the spectral range of highest Q-factors
too. For example, a 60~nm shift of the Q-band is observed for
$\mu$-disk of twice smaller diameter (Fig.~\ref{shift2}). In this
case, the only modification in $Q_{total}$ arises from a severe
attenuation of Q$_{rad}$'s, therefore, very low values of highest
$Q_{total}$ are observed. Moreover, the free spectral range in a
5~$\mu$m $\mu$-disk doubles with respect to a 10~$\mu$m one. We
underline, that the $\mu$-kylix configuration provides the same
amount of tuning of the Q-band with only a 0.4\% of effective
diameter modification, which also provides an almost unmodified FSR.

In conclusion, we have demonstrated a new class of bi-dimensional
micro-resonators with out-of-plane bending, which perfectly support
whispering-gallery modes. We refer to them as micro-kylixes. This
resonators are chip-integrable and can be easily fabricated using
standard microfabrication technology. Their particular geometry
reveals new physics: tuning of the highest Q-factors towards shorter
wavelengths, where the basic material has stronger absorption. This
is achieved through a smart interplay between radiative and material
quality factors. Our experimental results, validated by numerical
simulations, indicate that this physical phenomenon could be
exploited to obtain improved Q-factors in specific spectral windows
(shorter wavelength) without modifying resonator's physical size and
the free spectral range. Micro-kylix resonators can offer novel
technological solutions for micro-resonator physics and photonics
research and may open the door to new functionalities of resonator
devices, from sensing to optical amplification.\\

The authors thank Dr I. Carusotto for critical reading of the
manuscript. Helpful discussions with with P. Bettotti and M. Xie are
gratefully acknowledged.

\end{document}